\documentclass[12pt]{article}
\usepackage{amsmath,amssymb,cite}
\usepackage{graphicx}
\newcommand{\One}{{\hbox{{\rm 1{\hbox to 1.5pt{\hss\rm1}}}}}}

\newcommand{\vev}[1]{\langle\,#1\,\rangle}

\newcommand{\CB}{{\cal B}}

\newcommand{\CM}{{\cal M}}

\newcommand{\CO}{{\cal O}}

\newcommand{\CZ}{{\cal Z}}

\newcommand\be{\begin{equation}}
\newcommand\ee{\end{equation}}
\newcommand\bea{\begin{eqnarray}}
\newcommand\eea{\end{eqnarray}}
\newcommand\nn{\nonumber}
\newcommand\ba{\(\begin{array}}
\newcommand\ea{\end{array}\)}
\newcommand{\resection}[1]{\setcounter{equation}{0}\section{#1}}

\allowdisplaybreaks
\begin{document}

\setlength{\oddsidemargin}{0cm}
\setlength{\baselineskip}{7mm}

\begin{titlepage}
\renewcommand{\thefootnote}{\fnsymbol{footnote}}
%\begin{normalsize}
%\begin{flushright}
%\begin{tabular}{l}
%??? \\ ??? % 
%\end{tabular}
%\end{flushright}
% \end{normalsize} 
%  ~~~\\

\vspace*{0cm}
    \begin{Large}
%    \begin{bf}
       \begin{center}
         {Boundary correlation numbers in one matrix model}
       \end{center}
%    \end{bf}   
    \end{Large}
\vspace{0.7cm}

\begin{center}
Goro I{\sc shiki$^{1}$}\footnote
            {
e-mail address : 
ishiki@post.kek.jp}
    {\sc and}
Chaiho R{\sc im$^{2}$}\footnote
           {
e-mail address : 
rimpine@sogang.ac.kr}\\
      
\vspace{0.7cm}                    
{\it Department of Physics$^{2}$ and 
  Center for Quantum Spacetime (CQUeST)$^{1,2}$
}\\
{\it Sogang University, Seoul 121-742, Korea}
\end{center}

\vspace{0.7cm}

\begin{abstract}
\noindent

We introduce one matrix model coupled to multi-flavor vectors.
The two-flavor vector model is demonstrated
to reproduce the two-point correlation numbers 
of boundary primary fields of two 
dimensional $(2, 2p+1)$ minimal Liouville gravity on disk,
generalizing the loop operator (resolvent) description.
The model can properly describe non-trivial  
boundary conditions for the matter Cardy state  
as well as for the Liouville field.
From this we propose that 
the $n$-flavor vector model 
will be suited for producing the boundary correlation numbers
with $n$ different boundary conditions on disk. 

\end{abstract}
\vfill
\end{titlepage}
\vfil\eject

\setcounter{footnote}{0}

%\tableofcontents

%%%%%%%%%%%%%%%%%%%%%%%%%%%%%%%%%%%%%%%%%%%%%%%%%%%%%%%%%%%%%
\section{Introduction}
%%%%%%%%%%%%%%%%%%%%%%%%%%%%%%%%%%%%%%%%%%%%%%%%%%%%%%%%%%%%%
There have been developed two independent approaches to 
(Euclidean) two dimensional 
quantum gravity since the middle of 80's:
Liouville gravity (LG) \cite{Polyakov,KPZ,David,DK} 
versus matrix models (MM) \cite{BK,DS,GM,D,K,DFK}.  
Two approaches are checked in a number of particular models
\cite {S,GL,BAlZ}.  One may refer to more references in reviews in 
\cite{GM93,DFGZ}. 

Initiated by Al. Zamolodchikov's direct 
computation of partition function on fluctuating sphere 
at the simplest integral point 
(Liouville coupling set to $b^2 = \frac 25$)
\cite{zam_sphere},
Belavin and Zamolodchikov succeeded to 
confirm the correlation numbers of minimal gravity 
by providing so-called resonance transformation formula 
between conformal (Liouville) and KdV frame\cite{BZ}.  

Correlation numbers are given in terms of the integrated form of the local density (2-form operator) 
$\CO_k(X)$, $ O_k = \int_\CM~\CO_k(X)$ 
over the manifold $\CM$ which accommodates both ``matter" 
and the metric degrees of freedom localized at $X \in \CM$.
A generating function of the correlation
numbers in Liouville gravity is given by
$Z_{LG}(\{ \lambda_k\}) = \vev{e^{\sum_k \lambda_k O_k}}\,$.
A similar generating function,
which depends on the parameters $t_k$, can be introduced in Matrix models.
It was conjectured in \cite{MSS} 
that there exists a special choice of contact terms 
in MM or, equivalently, 
the special transformation $t_k = t_k (\{\lambda_j\})$
such that the coincidence of the partition functions 
is ensured in the $p$-critical one-matrix model OMM($p$) 
and the Minimal Liouville gravity MLG(2,2$p$+1)
for random surfaces with spherical topology.
This  relation between the parameters  $t_k$ and $\lambda_k$ 
was obtained in \cite{MSS} up to the  linear terms.
Its explicit form to all orders  was conjectured  in \cite{BZ} 
and checked up to the 4-th order \cite{BZ,GT}. 

Given the minimal gravity, it has been also shown 
\cite{BR} 
that the resonance transformation formula also works 
for bulk correlation numbers 
in the presence of fluctuating disk. 
In particular, the disk partition function of OMM($p$)
with the boundary length $l$ in the conformal frame
is given as \cite{MSS,BR}
\bea
\CZ_D(\mu, l)
 =\frac {u_0^{p+1}} {\sqrt l}  \, 
 \int_1^\infty  dx  L_p(x)  e^{- l u_0  x} 
 = \sqrt{\frac 2  \pi }\frac {u_0^{p+1/2}}{l}  K_{p+1/2}(u_0 l)
\eea
where $\mu$ is the bulk cosmological constant. 
On the other hand, the Liouville gravity partition function 
$Z_D(\mu,\mu_B ) $ is related 
by the inverse Laplace transform of $\CZ_D(\mu, l)$  
\be
\CZ_D (\mu, l) 
= l \int_{\uparrow} %0^ \infty 
\frac{d\mu_B}{2\pi i} ~ e^{\mu_B l}
~Z_D(\mu,\mu_B ) 
\,,
\label{diskpartition}
\ee
where the contour $\uparrow$ goes along the imaginary 
axis to the right from all the singularities 
of the integrand. 
Using the Laplace transform one can show that 
the resolvent in the continuum limit on the spherical topology
\be
\omega(z) 
= \left\langle {\rm tr}\;  \Big(\frac 1{z-M}\Big) \right\rangle 
=\left\langle
\int_0^{\infty}dl\; 
{\rm tr}\; e^{-l(z-M)}
\right\rangle
\ee
is proportional to $\cosh (\pi s/b)$ 
when the boundary parameter is set to $z=u_0 \cosh (\pi b s) $.
$u_0$ provides a scale parameter in the matrix model 
(details are given in appendix A).   
For $p$-critical theory, 
$\omega(z) $ has the scale dimension $u_0^{p+1/2}$ 
so that $\omega(z) = u_0^{p+1/2}\cosh (\pi s/b)$.

In Liouville field theory (LFT), 
$\,b$ corresponds to the Liouville coupling constant 
relating the background charge $Q= b+ 1/b $ 
and $z$ to the boundary cosmological constant 
$ \mu_B=\sqrt{{\mu}/{\sin(\pi b^2)}}
\cosh (\pi b s)$. 
MLG(2,2$p$+1) corresponds to the case $b^2 = 2/(2p+1)$. 
Boundary operator of LFT, 
$ ^{s_1}\left[e^{\beta \phi}(X)\right]^{\,\,s_2}$ 
with $X\in \partial \CM$ 
has conformal dimension 
$\Delta(\beta) = \beta (Q-\beta)$
and is specified by the boundary condition 
BC$(s_1;\, s_2)$ which $e^{\beta \phi}$ joins. 
The primary Liouville boundary operator 
in MLG(2,2$p$+1) is given as  
$B_{1\ell} = e^{\beta_{1\ell}\, \phi}\Phi_{1\ell}$ 
and $\beta_{1\ell} = b(1+\ell)/2$ with $\ell = 1, 2, \cdots, p$. 
$\Phi_{mn}$ represents the CFT matter in the Kac table
with the conformal dimension 
$\Delta_{mn} =  \alpha_{mn} (\alpha_{mn} -q)$ 
where $\alpha_{mn}= (n-1)b/2 - (m-1)/(2b)$ 
and $ q= 1/b -b$. 
It should be noted that in LFT,  
the boundary correlation of $e^{\beta \phi}$'s 
is known in \cite{FZZ, PT},
not to mention the boundary correlation 
for the A-series minimal model \cite{CL,Runkel}. 
However, the corresponding result is not  
available in the matrix model side, 
even though few attempts can be found 
in RSOS model and O(N) fluctuation models \cite{b-on-sos, b-on-sos2},
in loop gas model \cite{bloop,bloop2},
and in (two) matrix model \cite{Martinec:1991ht, b-two-matrix}. 

In this paper, we propose boundary changing operator 
description in OMM($p$) 
which reproduces boundary correlation numbers 
of primary fields 
in MLG(2,2$p$+1) on disk. 
This paper is organized as follows.
In section \ref{One matrix model with vectors},
we introduce one matrix model which couples to 
certain number of vectors. 
In section \ref{Correlation numbers},
we demonstrate that 2-point correlation number of the
integrated form of boundary changing operator,
$^{(s_1,1)}\CB_{1\ell}^{\,\,(s_2,\ell)}= \int_{\partial \CM}~^{(s_1,1)}B_{1\ell}^{\,\,(s_2,\ell)}(X)$ 
are reproduced from the 
matrix model with two vectors.
Here $s_1$ and $s_2$ correspond to 
the  boundary conditions of LFT and 
$1=(1,1)$ and $\ell=(1,\ell)$ 
are the Cardy label of minimal boundary conditions.
Starting from BC($s_1,1;\, s_2,\ell$) 
one can construct the general boundary condition, 
BC($s_1,m;\, s_2,n$),
when $1\le m,n \le p $ 
are allowed from fusion property.  
Section \ref{summary} 
is devoted to summary and discussion.

%%%%%%%%%%%%%%%%%%%%%%%%%%%%%%%%%%%%%%%%%%%%%%%%%%%%%%%%%%%%%%%%
\resection{One matrix model with vectors}
\label{One matrix model with vectors}
%%%%%%%%%%%%%%%%%%%%%%%%%%%%%%%%%%%%%%%%%%%%%%%%%%%%%%%%%%%%%%%%
In order to describe 2 dimensional gravity with boundaries,
we introduce one matrix model with vectors,
\begin{align}
e^Z=\int DM Dv^{(a)\dagger} Dv^{(b)} \exp
\left( -\frac{N}{g}{\rm tr}V(M) 
        -\sum_{a,b}v^{(a)\dagger}\cdot C^{(a,b)}(M) \cdot v^{(b)} \right),
\label{starting point}
\end{align}
where $M$ is a $N\times N$ Hermitian matrix
and $V(M)$ is a polynomial of $M$ 
which starts from a quadratic term $\frac{1}{2}M^2\,$.
$v^{(a)}$ 
and its hermitian conjugate $v^{(a)\dagger}$ 
are $N$ dimensional vectors 
and $\cdot$ represents the contraction of 
the $N$ dimensional indices. 
$a$ and $b$ label the ``flavors'' of the vectors. 
The number of flavors may depend on the number of different 
boundary conditions.
$C^{(a,b)}(M)$ is a hermitian matrix 
whose diagonal component is given 
as a polynomial of $M$, 
\begin{align}
C^{(a,a)}(M)=\sum_{n=0}^{K^a}b_n^{(a)} M^{K^a-n},
\label{CM}
\end{align}
with $b_0^{(a)}$ normalized\footnote{
Note that by rescaling the vectors, we can always 
fix the coefficient of the highest order term in $M$.
Nevertheless, to compare the matrix result with the Liouville gravity
one can choose a convenient normalization. 
More details are found in Sec.~\ref{Correlation numbers}.
}  
as $(-1)^{K_a}$
and $K^a$ is the order of the polynomial. 
$b_n^{(a)}$ in the diagonal component 
behaves as the source to the boundary preserving operator.
The off-diagonal component is also given as a polynomial 
in general and its coefficients are
the source to the boundary changing operator.

\begin{figure}[tbp]
\begin{center}
\includegraphics[width=0.8\textwidth, keepaspectratio, clip]{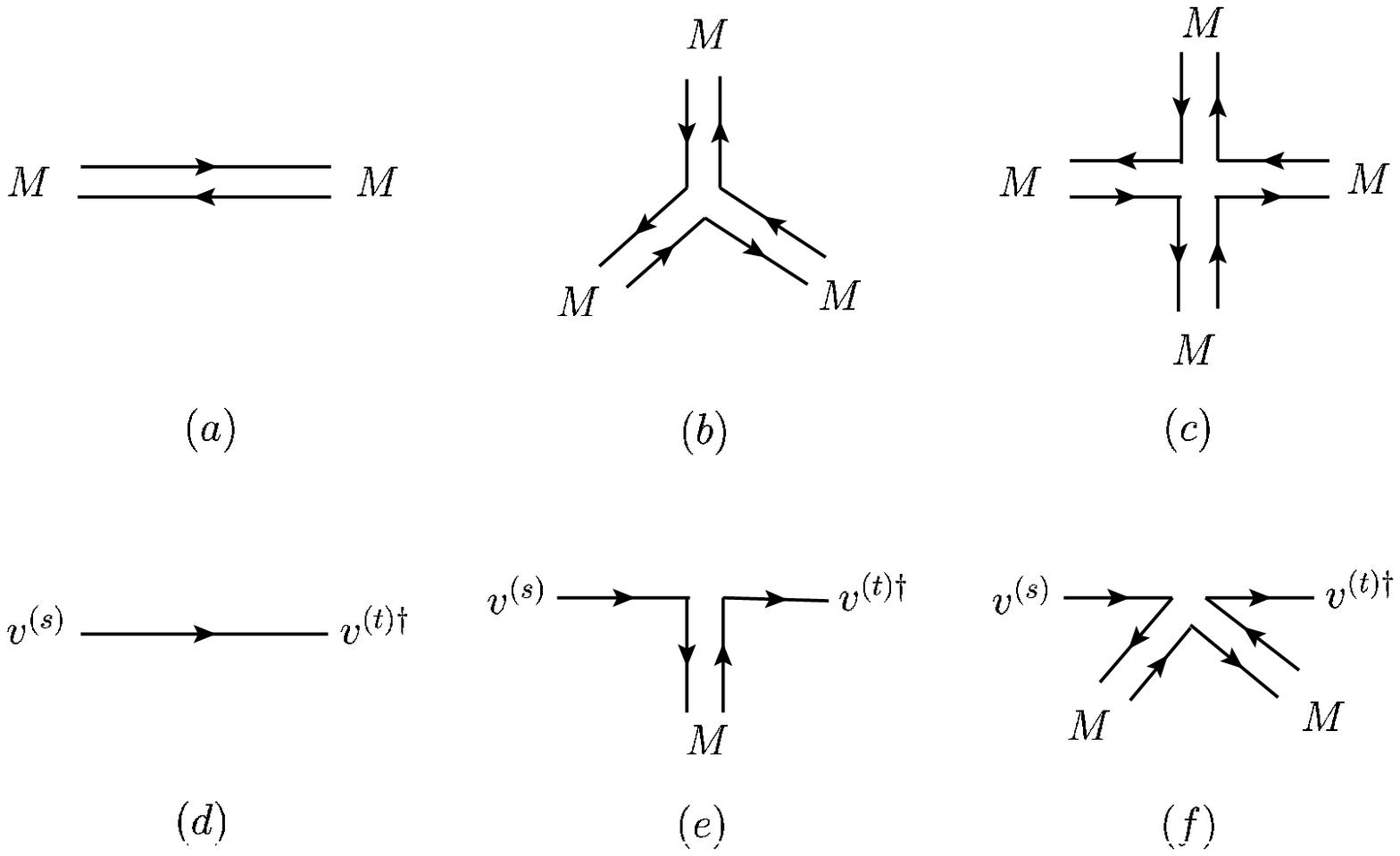}
\end{center}
\caption{(a): Propagator for $M$. (b)(c): Examples of bulk interactions.
(d): Propagator for the vectors. (e)(f): Examples of boundary interactions.}
\label{feynman diagram 1}
\end{figure}

\begin{figure}[tbp]
\begin{center}
%\hspace{-2cm}
\includegraphics[width=0.6\textwidth,keepaspectratio,clip]{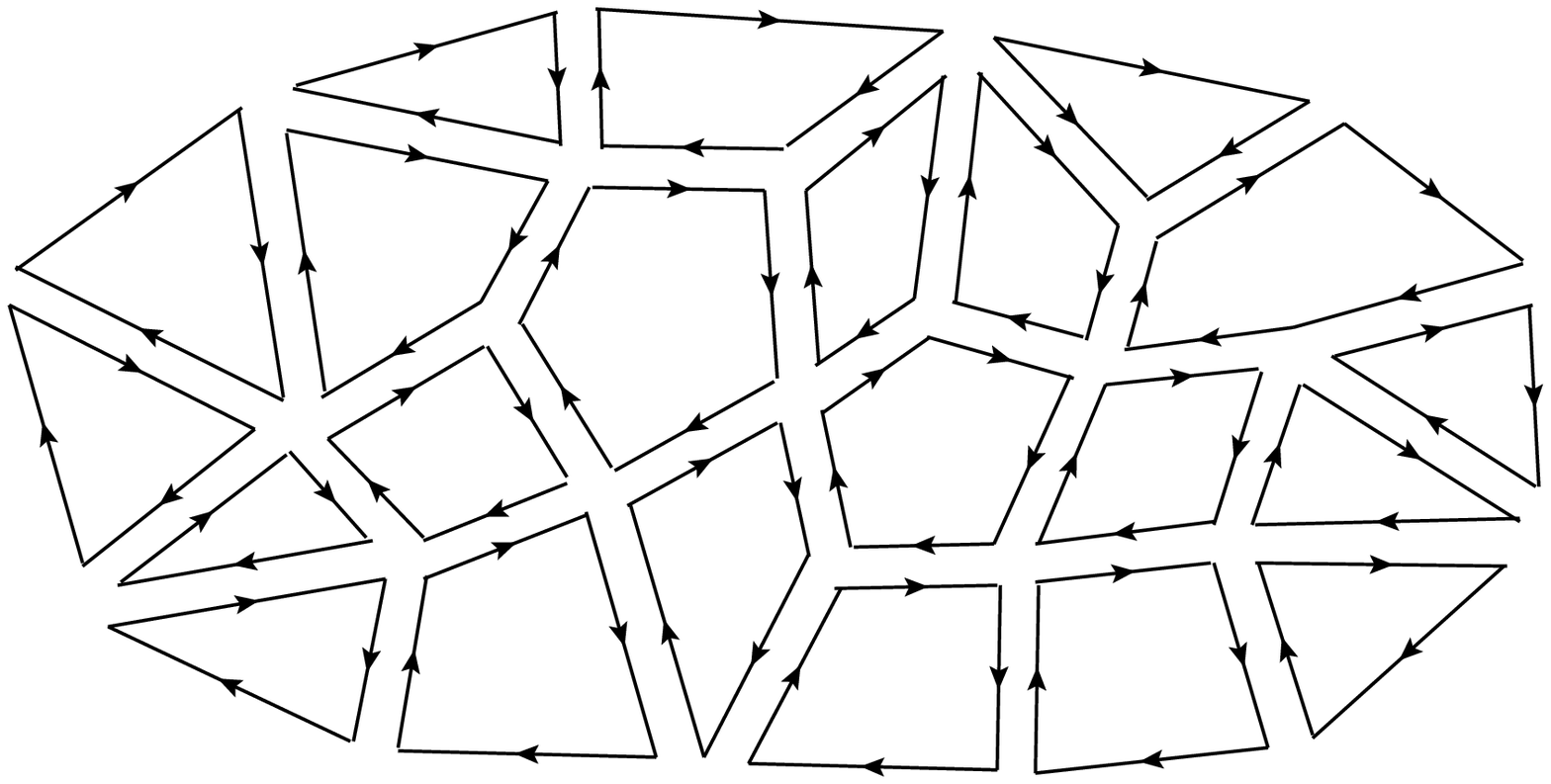}
\end{center}
\caption{Typical Feynman diagram with a boundary.}
\label{feynman diagram 2}
\end{figure}

This model can generate Feynman diagrams which correspond to
2 dimensional discretized surfaces.
We associate a double line with a propagator of $M$ and 
a single line with that of the vectors as usual.
The rule of drawing the Feynman diagram is shown in Fig. 
\ref{feynman diagram 1}.
A typical Feynman diagram
with a boundary is found in 
Fig. \ref{feynman diagram 2}.
Note that the vector model 
(\ref{starting point})
is quadratic in the vectors
and thus, the propagators of the vectors always form loops which
are regarded as boundaries.

The partition function $Z$ in (\ref{starting point}) 
can be expanded over various topologies 
including disks as following.
We first integrate out the vectors and obtain
\begin{align}
e^Z & =\int DM \exp \left( -\frac{N}{g}{\rm tr}V(M) 
  -{\rm Tr}\log C(M) \right) 
= e^{Z_{0}}\langle e^{-{\rm Tr}\log C(M)} \rangle_0, 
\label{integrate out vectors}
\end{align}
where ${\rm Tr}$ stands for the trace over both flavor
and matrix indices. $e^{Z_0}$ is the 
partition function without vectors
\begin{align}
e^{Z_0} = \int DM  \exp
\left( -\frac{N}{g}{\rm tr}V(M)  \right)
\label{partition function without boundary}
\end{align}
and describes manifolds without any boundary.
$\langle \cdots \rangle_0$ in 
(\ref{integrate out vectors}) stands for 
a normalized expectation value with respect to
(\ref{partition function without boundary}).
By taking the logarithm of 
(\ref{integrate out vectors}), we obtain
\begin{align}
Z& = Z_0 +\langle 
e^{-{\rm Tr}\log C(M)} \rangle_c
\equiv Z_0 + \sum_{h=1}^{\infty}Z_h,
\end{align}
where $\langle \cdots \rangle_c$
denotes the connected part of $\langle \cdots \rangle_0$.  
$Z_h$ becomes the partition 
function with $h \ge 1$ holes (boundaries) 
\begin{align}
Z_h=\frac{1}{h!} \langle (-{\rm Tr}\log C(M))^h  
\rangle_c
=\frac{1}{h!} \left\langle \left( \int_0^{\infty} \frac{dl}{l}
{\rm Tr}\; e^{-lC(M)}\right)^h  
\right\rangle_{\!\!\!c}. 
\label{partition function with h holes}
\end{align}

Noting that an insertion of a traced operator 
lowers the order of $N$ by $1$
and thus, $Z_h$ has the extra factor $N^{-h}$ 
compared with $Z_0$,
we can topologically expand 
$Z_h=\sum_{g=0}^{\infty}Z_h^g N^{\chi}$, where $\chi=
2-2g-h$ is the Euler characteristic for the manifold with
$g$ handles and $h$ holes. 
In the following, we restrict ourselves to the 
case of disk with no handle ($g=0,h=1$) for simplicity.

%%%%%%%%%%%%%%%%%%%%%%%%%%%%%%%%%%%%%%%%%%%%%%%%%%%%%%%%%%%%%%%%
\resection{Correlation numbers from one matrix model}
\label{Correlation numbers}
%%%%%%%%%%%%%%%%%%%%%%%%%%%%%%%%%%%%%%%%%%%%%%%%%%%%%%%%%%%%%%%%
 
Let us first consider the one-flavor case.  
Suppose $C(M)=x-M$. In this case,
(\ref{partition function with h holes}) with $h=1$
reduces to the well-known simple partition function on disk,
\begin{align}
Z_1= -\langle {\rm Tr} \log(x-M) \rangle_c
=\left\langle \int_0^{\infty} \frac{dl}{l}
{\rm Tr}\; e^{-l(x-M)}  
\right\rangle_{\!\!\!c} \,. 
\end{align}
The constant $x$ couples to the length of boundary
according to (\ref{diskpartition}) and 
is identified as the boundary cosmological 
constant, which is real.  Note that differentiating $Z_1$ 
with respect to $x$ gives the resolvent $\omega (x)$. 

Next, suppose the order of $C(M)$ is 2 ($K=2$). 
\begin{align}
C(M)=c_2 +c_1 M+M^2=(x^+-M)(x^--M)\,.
\end{align}
$c_1$ and $c_2$ are real and 
$ x^+ + x^- = -c_1$ and $ x^+ x^- = c_2$.
A certain correlation will be generated if one differentiates 
$Z_1$ with respect to the source $c_2$:
\begin{align}
O(x_1^+, x_1^-) 
\equiv 
\left\langle {\rm tr} \frac{1}{(x^+-M)(x^--M)}  \right\rangle 
= - \frac{w(x^+)-w(x^-)}{x^+-x^-}\,.
\label{1-point}
\end{align}
When $x^{\pm}$ are real, 
$O(x_1^+, x_1^-) $ is not vanishing 
and becomes the two point correlation number  
of the boundary cosmological operator 
$\CB_{11}$ \cite{b-two-matrix}. 
However, if $x^{\pm}$ is allowed complex
this is not the whole story
since $O(x_1^+, x_1^-) $ 
can vanish\footnote{This requirement is reminiscent
of the method of solving the resonance relation on 
sphere\cite{BZ}.}. 
In this case, $\partial Z_1 /\partial c_2$ 
can be interpreted as one point correlation 
of $\CB_{13}$ on disk.%
\footnote{(\ref{1-point}) is obtained from 
(\ref{b12}) when $x_1 \to \infty$
whose result is reproduced by fusing 
two $\CB_{12}$'s. We thank Bourgine for this point.
}
The vanishing condition is 
$\omega(x^{+}) = \omega(x^{-})$, 
or 
$s^{+}\pm  s^{-} =2ibn$
with $n$ integer.
In the following, the notations will be used 
$x^{(j)} = u_0\cosh(\pi b s^j)$
and 
$w(x^{(j)}) =  u_0^{p_0 +1/2} \, \cosh(\pi s^j/b)$. 

One can generalize the above argument to the case 
when $C(M)$ is the $\ell$-th order polynomial of $M$. 
$C(M)$ is factorized in terms of $\ell$ solutions,
$\{x^{(1)},\cdots, x^{(\ell)} \}$ satisfying $C(x^{(j)})=0$. 
Vanishing requirement on the one-point correlation  
$O(x^{(1)},\cdots, x^{(\ell)})$
gives a constraint between any of $x^{(j)}$'s
so that $\omega(x^{(j)})=\omega(x^{(k)})$ ($1\le j,k \le \ell$);
\begin{align} 
\frac{s^j \pm s^k} {2ib} = 0 \!\mod 1 \,.
%\label{s for vanishing one-point}
\end{align}

Next, two different boundary conditions need two flavors.
We put $C(M)$ as 
\begin{align}
C(M)= 
\left(
\begin{array}{cc}
x_1-M       & c^{12}   \\
c^{21}  & F(M)     \\
\end{array}
\right)
\end{align}
where $F(M)$ is a certain polynomial of $M$
and $c^{21}$ is a $M$-independent constant 
and is the complex conjugate of $c^{12}$. 
Differentiating 
the partition function $Z_1$
with respect to the sources $c^{12}$ and $c^{21}$
one has two point correlation of boundary changing 
operators 
\be
\left. 
\frac{\partial^2 Z_1}{\partial c^{12} \partial c^{21}}
\right|_{c^{ij}=0}
= \left\langle
{\rm tr}\left( \frac{1}{x_1-M}\frac{1}{F(M)} \right)
\right\rangle.
\label{two point function}
\ee

When $F(M)$ is linear in $M$, 
the correlation trivially reduces to 
that of $\CB_{11}$'s.
To describe non-trivial ones, 
we may put $ F(M)=(x_2^{(+)} -M)(x_2^{(-)} -M)$
and choose 
$s_2^\pm  = s_2 \pm  i b$ 
with $s_2$ real.
This choice ensures   
$\omega(x_2^{(+)}) =\omega(x_2^{(-)}) = -  \omega(x_2) $
with $x_2 = u_0\cosh (\pi b s_2 )$
and $C(M)$ hermitian. 
As the result, (\ref{two point function}) becomes
\bea
&&
O(x_1; \{x_2\}_2 )
\equiv 
\left\langle
{\rm tr}\frac1{(x_1-M)(x_2^{(+)} -M)(x_2^{(-)} -M) }
\right\rangle
\nn\\
&&\qquad =
 \frac{u_0^{p-\frac{3}{2}}\cosh\left(\frac{\pi s_p }{2b}   \right)
         \cosh\left(\frac{\pi s_m }{2b}   \right)}
          {2\sinh\left(\frac{\pi b (s_p+ib) }{2}   \right)
           \sinh\left(\frac{\pi b  (s_p-ib) }{2}   \right)
           \sinh\left(\frac{\pi b  (s_m+ib) }{2}   \right)
           \sinh\left(\frac{\pi b  (s_m-ib) }{2}   \right)}\,.
           \qquad
\label{b12}
\eea
where $s_p=s_1+s_2$ and $s_m=s_1 -s_2$. 
This exactly agrees with Liouville two-point correlation 
of $\CB_{12}$ with BC$(s_1;s_2)$ for the Liouville field.

%
%%%%%%%%%%%%%%%%%%%%%%%%%
\begin{figure}[btp]
\begin{center}
%\hspace{-2cm}
\includegraphics[width=0.8\textwidth, keepaspectratio, clip]{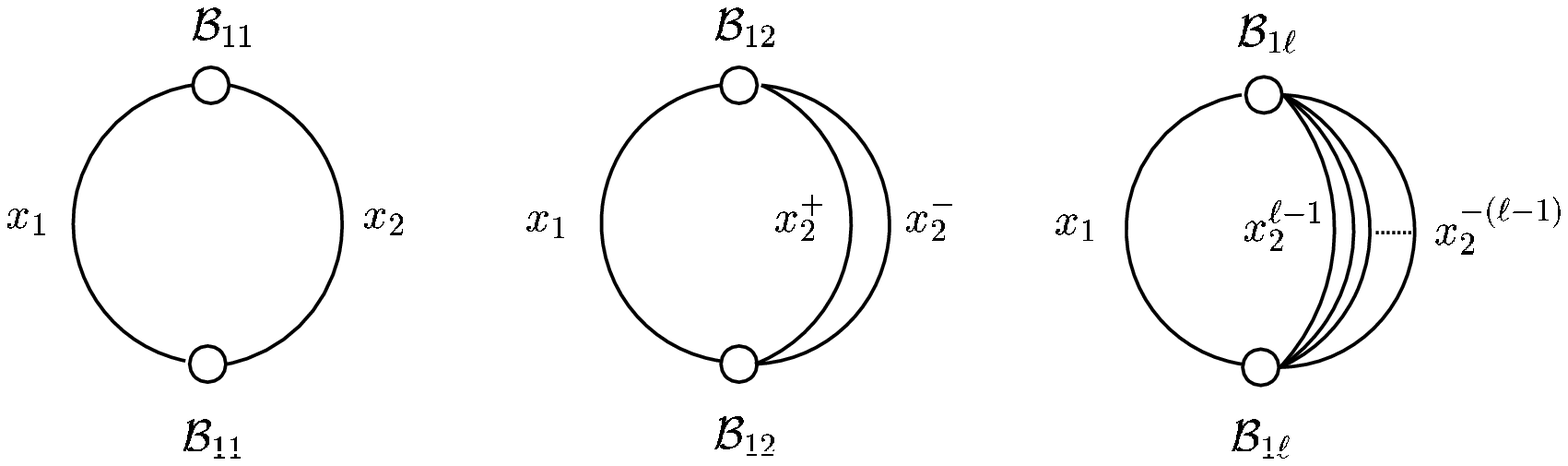}
\end{center}
\caption{The boundary 2-point correlations for ${\cal B}_{11}$,
${\cal B}_{12}$ and ${\cal B}_{1\ell}$. The line with $x$ 
represents the factor of $\frac{1}{x-M}$.}
\label{2-pt function}
\end{figure}
%%%%%%%%%%%%%%%%%%%%%%%%%
%

The generalization to other boundary operators is 
surprisingly simple: Put $F(M)$ as the $\ell$-th order polynomial  
and choose (with $s_2$ real)
\be
s_2^j  = s_2 +  i b j
\quad {\rm for}~~~j=-(\ell-1), -(\ell-3), \cdots, \ell-3, \ell-1 
\label{spm-ell}\,,
\ee 
so that $\omega(x_2^{(j)}) = \omega(x_2^{(j\pm2)})$. 
Note that the number of allowed complex parameters
for two point correlation is $\ell$ as 
shown in Fig.~\ref{2-pt function}.   
Explicit evaluation of 
(\ref{two point function}) is given as 
\be
O(x_1;  \{x_2\}_\ell )
\equiv\left\langle
{\rm tr}\left( \frac{1}{x_1-M}
\frac{1}{F_\ell(M;x_2)} \right)
\right\rangle
=(-)^{\ell} \frac{\omega(x_1)-\omega(x_2^{(\ell-1)})}
      {\prod_{k=0}^{\ell-1}(x_1 - x_2^{(\ell-1-2k)})} 
\qquad
\label{two-point-ell}
\ee
where $F_{\ell}(M;x)$ is the 
polynomials of $M$ with degree $\ell$, 
$\prod_{k=0}^{\ell-1} 
(x^{(\ell-1-2k)}-M)$.
This coincides with 
Liouville result of 
$^{s_1}\CB_{1\ell}^{s_2}$ 
up to BC-independent normalization\cite{FZZ}.
(See (A.16) of the first reference in \cite{bloop} 
for the explicit formula for $b^2 = 2/(2p+1)$.)  

One can also specify the boundary condition for the 
matter field in (\ref{two-point-ell}).
Noting the boundary structure constant 
$c^{(1,\ell,1)\, 11}_{1\ell,1\ell} =1$ \cite{Runkel},
one concludes that 
Eq.(\ref{two-point-ell}) produces 
$\left\langle \CB_{1\ell} \CB_{1\ell}  \right\rangle$
with BC($s_1,1;\, s_2,\ell$) and $F_\ell(M;x)$ generates 
$\ell=(1,\ell)$ Cardy boundary condition.
This conclusion is backed up by 
one-point correlation of $\CB_{11}$ 
with BC($s,k$), 
\be
O(\{x\}_k) = \frac{\partial }{\partial x }
\mbox{\rm tr} \log(F_k (M;x)) =[k]_q \,\omega(x) 
\label{1-pt-bc}
\ee
where $[x]_q = {(q^x-q^{-x})}/{(q-q^{-1})}$ 
is the $q-$number with $q= \exp(i \pi b^2)$. 
From the field theoretic point of view, 
one can put the ratio
$ {O(\{x\}_k)}/{O(\{x\}_1)}$
as the ratio of the vacuum expectation value 
\cite{Runkel} of 
the matter part  $ {\left\langle 1\right\rangle_k}/
{\left\langle 1\right\rangle_1} =[k]_q$
since Liouville contribution cancels out 
in the ratio.
The result is consistent with Eq.~(\ref{1-pt-bc}).

Incidentally, we note that 
the matter operator is identified as 
$\Phi_{1,k}= \Phi_{1, 2p+1-k}$
and BC($s;k$)=BC($s;1,2p+1-k$)
and $\CB_{1,k}$ is related 
to  $\CB_{1, 2p+1-k}$ 
through the Liouville reflection
(``unitary'' condition)
\cite{FZZ,bloop}. 
Since the matrix result Eq.~(\ref{two-point-ell}) 
shows the same functional dependence as 
the Liouville result,  
it goes without saying that 
the same Liouville reflection holds for
the matrix result \cite{bloop}.
Thus, even though the boundary primary operator 
$\CB_{1,k}$ is allowed as $(1\le k \le 2p)$ 
and the order of the diagonal component 
of $C(M)$ in Eq.~(\ref{CM}) can be restricted to 
$1\le K^a \le 2p$ for the $p$-critical model,
the number of independent operators is further reduced 
by half by the Liouville reflection.

From the above consideration, we assert that
$O(\{x_1\}_\ell; \{x_2\}_m)$
describes the two-point correlation 
of $\CB_{1,\ell+m-1}$ with BC($s_1,\ell;\,s_2,m$).
One can support this idea using the fusion rule.
Let us consider the four-point correlation
with simple BC 
($\ell \le m$ for definiteness)
\be
\left\langle 
^{(s_1;1)}\left[\CB_{1\ell}\right]^{(s_2;\ell)}\,
^{(s_2;\ell)}\left[\CB_{1\ell}\right]^{(s_3;1)}\,
^{(s_3;1)}\left[\CB_{1m}\right]^{(s_4;m)} \,
^{(s_4;m)}\left[\CB_{1m}\right]^{(s_1;1)} 
\right\rangle \,.
\label{4pt-LG}
\ee
If one uses the operator fusion rule
\bea
 ^{(s_4;m)}\left[\CB_{1m}\right]^{(s_1;1)} \otimes \,
^{(s_1;1)}\left[\CB_{1\ell}\right]^{(s_2;\ell)}  
&= \oplus_{\left|m-\ell+1 \right| \le k \le m+\ell-1 }\,
 ^{(s_4;m)}\left[\CB_{1k}\right]^{(s_2;\ell)} 
\nn \\
^{(s_2;\ell)}\left[\CB_{1\ell}\right]^{(s_3;1)} \otimes\,
^{(s_3;1)}\left[\CB_{1m}\right]^{(s_4;m)} 
&= \oplus_{\left|m-\ell+1 \right| \le k \le m+\ell-1 }\,
 ^{(s_2;\ell)}\left[\CB_{1k}\right]^{(s_4;m)} 
 \label{elementary-fusion}
\eea
one ends up with the two point correlation   
with general BC's.
\be
 \left\langle 
^{(s_2;\ell)}\left[\CB_{1k}\right]^{(s_4;m)} \,
^{(s_4;m)}\left[\CB_{1k}\right]^{(s_1;\ell)} 
\right\rangle \,.
\label{dressed-2pt}
\ee
 
From the matrix point of view, the four point correlation 
can be conjectured of the form  
\be
\left\langle
{\rm tr}\left( 
\frac{1}{F_1(M;x_1)}\,
\frac{1}{F_\ell(M;x_2)} \,
\frac{1}{F_1(M;x_3)} 
\frac{1}{F_m(M;x_4)} \right)
\right\rangle\,.
\label{4pt-matrix}
\ee 
Two point correlation is obtained 
by contracting the 1-boundary parts,
which expands $1/F_1(M;x)$ in powers of $M/x$:%
\footnote{
$M/x$-expansion 
is equivalent to the small length 
expansion if one uses the Laplace 
transform Eq.~(\ref{diskpartition}).
}
\be
\left\langle
{\rm tr}\left( M^g
\frac{1}{F_\ell(M;x_2)} 
\frac{1}{F_m(M;x_4)} \right)
\right\rangle
\label{matrix-2pt}
\ee 
with $0 \le g \le \ell-1$.\footnote{  
When $g \ge \ell$, 
Eq.~(\ref{matrix-2pt}) reduces 
to the summation of Eq.~(\ref{matrix-2pt})'s 
with $g <\ell$ except $\ell=m=1$.
} 
This shows that the 
number of allowed correlations
is the same as that of Eq.~(\ref{dressed-2pt})
and the range of $k$ is  
$\ell+m-1 -2g\le k \le l+m-1$.
It is noted that (\ref{matrix-2pt}) with $g\neq 0$ can be 
generated by $C(M)$ with $M$-dependent off-diagonal blocks,
\begin{align}
C(M)= 
\left(
\begin{array}{cc}
F_{\ell}(M;x_1)    & G(M)   \\
G(M)^{\dagger}  & F_{m}(M;x_2)   \\
\end{array}
\right).
\label{2 polynomials}
\end{align}

When $g=0$, Eq.~(\ref{matrix-2pt})
is indeed the 2-point correlation 
of ${\cal B}_{1,\ell+m-1}$
with BC($s_1,\ell;\, s_2, m $) up to 
a normalization.
To elaborate on this, let us consider
the two-flavor matrix model
of the form (\ref{2 polynomials}) with $G(M)$ given by 
a $M$-independent constant: $G(M)=c^{12}$.
For $\ell=k=2$,  
one finds a relation after a short calculation,
\bea
\left.
\frac {\partial^2 Z_1}{\partial c^{12}\,\partial c^{21}} 
\right|_{c^{ij}=0} = 
 O( \{x_1\}_2; \{x_2\}_2 )
=2 \cos(\pi b^2) \,
O(x_1; \{x_2\}_3 )\,.
\eea 
This relation connects
BC($s_1,2;\, s_2, 2 $) 
and 
BC($s_1,1;\, s_2, 3 $)
for the correlation of $\CB_{13}$. 
One can show a general recursion relation 
between the two-point correlations of ${\cal B}_{1,\ell+m-1}$ with
different boundary conditions: 
\be
[m-1]_q \,\,  
O\left( \{x_1\}_{\ell} ; \{x_2\}_{m} \right)
=[\ell]_q \,\,
O\left(\{x_1\}_{\ell+1}; \{x_2\}_{m-1} \right)\,.
\label{2pt-recursion}
\ee
Thus, the matrix model predicts the obvious connection between 
BC($s_1,a; s_2,b)$ 
with $a+b=\ell+m$
for the correlation of $\CB_{1, \ell+m-1}$.

One can demonstrate 
that (\ref{2pt-recursion}) is consistent with
the field theoretic result.
For the matrix side,  one can rescale 
the matrix $F_k (M;x) $ so that it has the form, 
$N_k \, F_k (M;x) $ 
where $N_k$ is a constant.  
Then, the two point correlation 
is rescaled as
\be
\frac1{N_k \, N_{\ell+1} }\,\,
O\left( \{x_1\}_k ; \{x_2\}_{\ell+1} \right)\,.
\label{M-corr}
\ee
For the field theoretic side, 
one considers the operator 
$\Lambda_{1,k+\ell}\,\,
e^{ \beta_{1,k+\ell} \, \phi} \, \Phi_{1,k+\ell}$,
introducing the normalization constant $\Lambda_{1, k+\ell}$  
independent of the BC. 
The two point correlation 
of $\CB_{1, k+\ell}$ 
with BC($s_1,k; s_2, \ell+1$)  
is given as  
\be
\Lambda_{1,k+\ell}^2\,
d_L ( \beta_{1, k+\ell}, s_1, s_2)\,\,
d_M(k;\ell+1)\,.
\label{F-corr}
\ee
Here, $d_L$ is the Liouville correlation 
and $d_M$ is the matter part correlation.
Equating two results, 
Eq.~(\ref{M-corr}) and Eq.~(\ref{F-corr}) 
one has 
\be
O\left( \{x_1\}_k ; \{x_2\}_{\ell+1} \right)
= {N_k \, N_{\ell+1} }  \, 
\Lambda_{1,k+\ell}^2\,
d_L ( \beta_{1, k+\ell}, s_1, s_2)\,\,
d_M (k;\ell+1)\,.
\label{matrix-field}
\ee 

On the other hand, 
Eq.~(\ref{matrix-field}) should be compatible 
with Eq.~(\ref{2pt-recursion}),
which leads to nontrivial relations 
between $N_k$'s.
$
N_k \, N_{\ell+1} \, d_M (k,\ell+1) 
= 
N_{k+1} \, N_{\ell} \,\, d_M (k+1,\ell)\,
$.
(Note that Liouville part cancels out 
because the same Liouville BC($s_1, s_2$) and
the same operator $\CB_{1, k+\ell}$ are used in both sides.)
$d_M$ is given in terms of boundary structure constant \cite{Runkel}
\be
d_M (k;\ell+1) 
= c^{(k, \ell+1, k)\,(11) }_{(1,k+\ell),(1,k+\ell)} \, 
\left\langle 1 \right\rangle_k
= c^{( \ell+1, k, \ell+1)\,(11) }_{(1,k+\ell),(1,k+\ell)} \, 
\left\langle 1 \right\rangle_{\ell+1} \,.
\ee
This simplifies the relation as   
\be
N_k \, N_{\ell+1} \, \,c^{(k, \ell+1, k)\,(11) }_{(1,k+\ell),(1,k+\ell)} 
= 
N_{k+1} \, N_{\ell} \,\, c^{(\ell, k+1, \ell)\,(11) }_{(1,k+\ell),(1,k+\ell)} .
\label{N-recursion}
\ee
One can find a consistent solution of $N_k$'s to Eq.~(\ref{N-recursion})
which is crucial for the consistency of (\ref{matrix-field}).
Considering the boundary  structure constant 
is identified with the fusion matrix \cite{Runkel}
$
c^{(A,B,C)\,(K) }_{LM} = F_{B,K}
\left[
\begin{array}{l}
A,\, C \\ L,\,M  
\end{array}
\right] $,
and the fusion matrix
is written in a factorized form
when $K=(11)$ 
\cite{MS-lecture},
$
F_{(1i) ,(11)} 
\left[
\begin{array}{l}
(1k),\, (1k) \\ (1j),\,(1j)  
\end{array}
\right] 
= \sqrt{\frac{ F_j F_k }{F_i }}
$
with $F_i = 1/[i]_q$, 
one has
\be
\frac{ N_{k+1}} {N_k} \sqrt{ \frac 1 {F_k F_{k+1}}} 
= \frac{N_{\ell+1}} {N_\ell} \sqrt {\frac1 {F_\ell F_{\ell+1}}} \,.
\label{recursion for N_k}
\ee 
Thus, $N_k (k\ge 3)$ is determined completely 
from $N_1$ and $N_2$. 
Finally, putting $k=1$ in Eq.~(\ref{matrix-field}),
one finds the field normalization $\Lambda_{1,\ell}^2$
in terms of $N_1\,N_{\ell}$. 
Once $N_k$'s are determined by the recursion relation
(\ref{recursion for N_k}),
the identification (\ref{matrix-field}) is consistent
with the relation (\ref{2pt-recursion}) found in the matrix model.

%%%%%%%%%%%%%%%%%%%%%%%%%%%%%%%%%%%%%%%%%%%%%%%%%%%%%%%%%%%%%%%%
\resection{Summary and discussion}
\label{summary}
%%%%%%%%%%%%%%%%%%%%%%%%%%%%%%%%%%%%%%%%%%%%%%%%%%%%%%%%%%%%%%%%

We propose a generalized partition function 
of one-matrix model (\ref{integrate out vectors})
to give boundary correlation numbers of 
primary fields of MLG(2,2$p$+1)
on disk. 
We demonstrate explicitly that the
two-flavor vector model 
correctly reproduces  
the two-point correlation numbers 
of boundary primary operator 
$\CB_{1,k}\, (1\le k \le 2p)$ in MLG(2,2$p$+1).

A few comments follow. 
First, one can obtain the general boundary condition
by contracting 1-boundary on the disk
as given in 
Eq.~(\ref{dressed-2pt}) and 
Eq.~(\ref{matrix-2pt}).
One may equally contract 
$\ell$ and $m$-boundaries
in Eq.~(\ref{4pt-LG}). 
In this case one is left with 
\bea
^{(s_1;1)}\left[\CB_{1\ell}\right]^{(s_2;\ell)}  
\otimes \,
^{(s_2;\ell)}\left[\CB_{1\ell}\right]^{(s_3;1)}
&= \,^{(s_1;1)}\left[\CB_{11}\right]^{(s_3;1)} 
\\
^{(s_3;1)}\left[\CB_{1m}\right]^{(s_4;m)} 
\otimes\,
^{(s_4;m)}\left[\CB_{1m}\right]^{(s_1;1)} 
&=\, ^{(s_2;1)}\left[\CB_{11}\right]^{(s_1;1)} 
\nn
\eea 
since $(1,1)$-boundary allows only (1,1) operator. 
This property is seen in the matrix side  
by contracting 
$\ell$ and $m$-boundaries
in Eq.~(\ref{4pt-matrix}).  
Eq.~(\ref{matrix-2pt}) 
is left with the two-point correlation of $F_1$'s only 
\be
\left\langle
{\rm tr}\left( 
\frac{1}{F_1(M;x_1)} 
\frac{1}{F_1(M;x_3)} \right)
\right\rangle
\ee  
and its contribution is 
$\CB_{11}$ with  BC($s_1, 1; s_2,1$). 
 
Second, the non-vanishing power of $M$ ($g\ne 0$) 
in Eq.~(\ref{matrix-2pt}) 
can be obtained using $C(M)$ in Eq.~(\ref{CM}) 
when 
its off-diagonal component
contains the $M$ dependent term up to the power $g$
which is less than any degree of the corresponding 
diagonal component. 
Therefore, the $M$-dependent off-diagonal term 
produces the two-point correlation of $\CB_{1,k}$ of 
BC($s_1, a; s_2,b$) with $a+b \ne k+1$. 
However, it is not clear yet,
what kind of mechanism restricts the maximum power $g$.
In addition, fusing rule is not simple for 
non-vanishing $g$ as shown in Eq.~(\ref{matrix-2pt}).
This suggests that one needs to find the more 
detailed description of 
the $M$-dependency of the off-diagonal term.

Finally, is the proposal working for three point correlation 
with 3-flavor vectors? 
Let us consider $O(x_1; x_2; x_3) $. 
This produces the 3-point correlation
of $\CB_{11}$'s in \cite{b-two-matrix}. 
On the other hand, 
the Liouville boundary 3-point correlation 
${\cal C}_{\beta_1\beta_2\beta_3}^{(s_2, s_3, s_1)}$
of $e^{\beta_i \phi}$'s  
with BC$(s_1;\, s_2;\, s_3)$ 
satisfies the difference relation%
\footnote{We fix a typo in the original equation.}
\cite{PT,b-on-sos}
\bea
&-\sinh\left\{\frac{\pi b}{2}(s_2-s_1+ib-2i\beta_3) \right\}
\sinh\left\{\frac{\pi b}{2}(s_3-s_1+ib-2i\beta_2) \right\}
C_{\beta_1\beta_2\beta_3}^{(s_2, s_3, s_1+ib)}
\nonumber\\
&+\sinh\left\{\frac{\pi b}{2}(s_2+s_1+ib+2i\beta_3) \right\}
\sinh\left\{\frac{\pi b}{2}(s_3+s_1+ib+2i\beta_2) \right\}
C_{\beta_1\beta_2\beta_3}^{(s_2, s_3, s_1-ib)}
\nonumber\\
&=\sinh\left\{\frac{\pi b}{2}(s_2+s_3+4ib-2i\beta_1) \right\}
\sinh (\pi b s_1)
C_{\beta_1\beta_2\beta_3}^{(s_2+ib, s_3+ib, s_1)}.
\label{difference eq}
\eea
$O(x_1; x_2; x_3) $ obviously satisfies (\ref{difference eq}) 
with $\beta_1=\beta_2=\beta_3=b$. 
One can consider a more general form. 
For example, 
$O(x_1; x_2; \{x_3\}_2) $ 
satisfies 
the difference equation (\ref{difference eq}) with 
$(\beta_1,\beta_2,\beta_3)=(b,\frac{3b}{2},\frac{3b}{2})$
and corresponds to 
$\langle {\cal B}_{11}{\cal B}_{12}{\cal B}_{12} \rangle$
with BC$(s_2, 2;\, s_3,1;\, s_1,1)$. 
Likewise, we expect that 
the $n$-point correlation number
$ O(x_1; \cdots ; x_n)$ 
will produce the result
$\left\langle \CB_{11} \cdots  \CB_{11} \right\rangle$ 
with BC$(s_1,1;\, \cdots;\, s_n,1)$. 
Nevertheless, we need more careful check for 
correlations of other non-trivial operators of $\CB_{1,\ell}$'s.
In addition, it looks very plausible that the same 
idea can work for the boundary correlation 
of multi-matrix model.
Further details will be reported elsewhere in the near future. 

{\sl Note added:} After completion of this work  authors of 
\cite{BBR} let us know that the same model arises in 
string theory on partially resolved singularities.

%%%%%%%%%%%%%%%%%%%%%%%%%%%%%%
\section*{Acknowledgements}
We are indebted to A. Belavin and J.-E. Bourgine
for the discussion at the initiation of this work
and to I. Kostov for encouraging remarks. 
The work is partially supported 
by the National Research Foundation of Korea (KNRF) 
grant funded by the Korea government (MEST)
2005-0049409 (I \& R) and R01-2008-000-21026-0 (R). 

\appendix
%%%%%%%%%%%%%%%%%%%%%%%%%%%%%%%%%%%%%%%%%%%%%%%%%%%%%%%%%%%%%
\resection{Resolvent in the spherical limit}
%%%%%%%%%%%%%%%%%%%%%%%%%%%%%%%%%%%%%%%%%%%%%%%%%%%%%%%%%%%%%
We present an expectation value of the resolvent
\begin{align}
\omega(z) = 
\left\langle
{\rm tr}\; \frac{1}{z-M}
\right\rangle
=\left\langle
\int_0^{\infty}dl\; 
{\rm tr}\; e^{-l(z-M)}
\right\rangle
\end{align}
in the large-$N$ limit.
In the double scaling limit (continuum limit), 
$M \sim a_1-a_2Q\,$ 
where $a_1, a_2$ are constants
and $Q=\frac{d^2}{dx^2}+u(x)$
\cite{GM, MSS,Banks}. 
Thus, one may renormalize  $ z \to a_1 +a_2 z$
and put 
\be
 z=u_0 \cosh (\pi b s)
\ee 

In the large-$N$ limit,  $u(x)$ 
is given by a maximal real solution to the string equation
\begin{align}
P(u,\{t_k\})=u^{p+1}+\sum_{k=0}^{p-1}t_k u^{p-k-1}=0,
\end{align}
with $t_{p-1}=x$.  
The parameter $-t_0$ is proportional to the bulk
cosmological constant $\mu$ and 
the other parameters, $\{t_k|1 \leq k\leq p-1 \}$, 
describe the relevant deviations
from the $p$-critical point in the KdV frame. They 
are related, through the so-called resonance 
transformation, to the perturbative coupling constants
${\lambda_k}$ which couple to 
the operators in the minimal gravity.
The resonance relation was solved in
\cite{BZ} and 
$P(u,\{t_k \})$ is given in terms of $\{ \lambda_k \}$ as follows,
\begin{align}
P(u,\{t_k(\lambda_k)\})= u_0^{p+1}\, \frac{(p+1)!}{(2p-1)!!}
\left(
\frac{L_{p+1}(u/u_0)
-L_{p-1}(u/u_0)}{2p+1}\right)
+{\cal O}(\lambda_k),
\end{align}
where 
%$g=(p+1)!/(2p+1)!!$, 
$u_0$ is a solution to 
the string equation with $t_k=0$ for $k  \geq 1 $, namely,
$u_0=\sqrt{-t_0}\sim \sqrt{\mu}$ and 
$L_p(x)$ is the Legendre polynomial.

In the large-$N$ limit, we can neglect 
the commutator $[\frac{d}{dx}, u(x)]=0$, 
\begin{align}
\omega(z) 
&=\left.
\int_0^{\infty}\frac{dl}{2a_2 \sqrt{\pi l}} 
\int_{t_0}^{\infty} dx e^{-l(z-u(x))}
\right|_{\lambda_k=0}
\nonumber\\
&=\left.
\int_0^{\infty}\frac{dl}{2a_2 \sqrt{\pi \ell~}} 
\int_{u(x=t_0)}^{\infty} du \frac{dP(u)}{du} e^{-\ell(z-u)}
\right|_{\lambda_k=0}
\nonumber\\
&=u_0^{p+1}\, \frac{(p+1)!}{2(2p-1)!!a_2}
\int_0^{\infty}\frac{dl}{\sqrt{\pi \ell~}} 
e^{-lz}
\int_1^{\infty} dy L_{p}(y) e^{-lu_0y}.
\end{align}
If one uses
\bea
\int_1^{\infty} dx L_n(x)e^{-lx} 
&=&\sqrt{\frac{2}{\pi \ell}\,}\, K_{p+1/2}(l)
\nn\\
\int_0^{\infty} \frac{dl}{2\pi l}e^{-l\cosh(s)} K_{p+\frac{1}{2}}(l)
&=& \frac{(-1)^{p+2}}{2p+1} \, 
\cosh \left( (p + 1/2)s \right)
\nn
\eea 
for the Macdonald function of a half-integer
order $K_{p+1/2}(x)$, 
one obtains the singular part of the resolvent in the 
large-$N$ limit,
\be
\omega(z) =\alpha u_0^{p+\frac{1}{2}}
\cosh\left(\frac{\pi s}{b} \right)
\ee
where $\alpha  = 
\frac{(-1)^{p+2}\sqrt{2}(p+1)!} { a_2 (2p+1)!!}$ 
is  independent of $s$
and can be absorbed in the renormalization.

%%%%%%%%%%%%%%%%%%%%%%%%%%%%%%%%%%%%


\begin{thebibliography}{99}
%%%%%%%%%%%%%%%%%%%%%%%%%%%%%%%%%%
\raggedright
\parskip 1pt
%
%
\bibitem{Polyakov}
A.\ Polyakov, 
%``Quantum geometry of bosinic strings'', 
Phys.\ Lett.\ {\bf B103} (1981) 207.
%
\bibitem{KPZ}
V.\ Knizhnik, A.\ Polyakov and A.\ Zamolodchikov,
%``Fractal Structure of 2D quantum gravity'',
Mod.\ Phys.\ Lett.\  {\bf A3} (1988) 819. 
%
\bibitem{David}
F.\ David,
%``Conformal Field Theories Coupled to 2-D Gravity in the Conformal Gauge'', 
 Mod.\ Phys.\ Lett.\  {\bf A3} (1988) 1651.
%
\bibitem{DK}
J.\ Distler and H.\ Kawai,   
%``Conformal Field Theory and 2D Quantum Gravity Or Who's Afraid of Joseph 
%Liouville?'', 
Nucl.\ Phys.\ {\bf B321} (1989) 509.
%
\bibitem{BK}
E.\ Brezin and V.\ Kazakov,
%``Exactly Solvable Field Theories Of Closed Strings'',
Phys.\ Lett.\ {\bf B236} (1990) 144. 
%
\bibitem{DS}
M.\ Douglas and S.\ Shenker,
%``Strings in Less Than One-Dimension'', 
Nucl.\ Phys.\ {\bf B335} (1990) 635.
%
\bibitem{GM}
D.\ Gross and A.\ Migdal,
%``Nonperturbative Two-Dimensional Quantum Gravity'',
 Phys.\ Rev.\ Lett.\ {\bf 64} (1990) 127;
% ``A Nonperturbative Treatment Of Two-Dimensional Quantum Gravity'',
Nucl.~Phys.~{\bf B340} (1990) 333. 
%
\bibitem{D}
M.\ Douglas,
%``Strings In Less Than One-Dimension And The Generalized K-D-V Hierarchies'', 
Phys.~Lett.~{\bf B238} (1990) 176. 
%
\bibitem{K}
V.\ Kazakov,
%``The Appearance of Matter Fields from Quantum Fluctuations of 2D Gravity'',
Mod.\ Phys.\ Lett {\bf A4} (1989) 2125.
%
\bibitem{DFK}
P.\ Di Francesco and D.\ Kutasov,
%``World sheet and space time physics in two 
%dimensional (super) string theory'',
Nucl.~Phys.~{\bf B375} (1992) 119.
%
\bibitem{S}
M.\ Staudacher,
%``The Yang-Lee singularity on a dynamical planar random surface'',
Nucl.~Phys.~{\bf B336} (1990) 349. 

%
\bibitem{GL}
M.\ Goulian and M.\ Li,
%``Correlation functions in Liouville theory'',
Phys.~Rev.~Lett.~{\bf 66} (1991) 2051.
%

\bibitem{BAlZ}
A.\ Belavin and Al.\ Zamolodchikov,
%``Moduli integrals,ground ring and four-point 
%function in minimal Liouville gravity'',
Theor.~Math.~Phys.~{\bf  147} (2006) 729.
%{\tt arXiv:hep-th/0510214}

\bibitem{GM93}
P.\ Ginsparg and G.\ Moore,
``Lectures on 2-D gravity and 2-D string theory (TASI 1992)'',
{ arXiv:hep-th/9304011}
%
\bibitem{DFGZ}
P.\ Di Francesco, P.\ Ginsparg and J.\ Zinn-Justin,
%``2-D gravity and random matrices'',
 Phys.~Rep.~{\bf 254} (1995) 1.  
%

\bibitem{zam_sphere}
 Al. Zamolodchikov,  
``Perturbed Conformal Field Theory on Fluctuating Sphere'',
arXiv:hep-th/0508044 (2005). 
%
\bibitem{BZ}
A.A.\ Belavin and A.\ Zamolodchikov, 
%``On correlation numbers in 2D minimal gravity and matrix models'',
J.~Phys.~{\bf A42} (2009) 304004.
%{\tt arXiv: 0811.0450[hep-th]}  .
%
\bibitem{MSS}
G.~Moore, N.~Seiberg, M.~Staudacher, 
%``From loop to states in 2D quantum gravity'',
Nucl.~Phys.~{\bf B362} (1991) 665.
%
\bibitem{GT}
G.\ Tarnopolsky,
``Five-point Correlation Numbers in One-Matrix Model'',
{ arXiv: 0912.4971[hep-th]}
%
\bibitem{BR} 
A. Belavin and C. Rim,
%``Bulk one-point function on disk in one-matrix model'', 
Phys.\ Lett.\ {\bf B687} (2010) 264. 

\bibitem{FZZ}
V.\ Fateev, A.\ Zamolodchikov and  Al.\ Zamolodchikov, 
``Boundary Liouville field theory. 1. Boundary 
state and boundary two point function'', 
{ hep-th/0001012}. 

\bibitem{CL}
J.\ L.\ Cardy, Nucl.\ Phys.\ {\bf b 240} (1984) 514;
 J.\ L.\ Cardy and D.\ C.\ Lewellen, Phys.\ Lett.\ 
 {\bf B259} (1991) 274.  

\bibitem{Runkel}
I.\ Runkel, 
%``Bounday structure constants for the A-series Virasoro minimal model'',
Nucl.\ Phys.\ {\bf b 549}[FS] (1999) 563. 


\bibitem{PT}
 B.~Ponsot, J.~Teschner, 
%``Boundary Liouville Field Theory: Boundary Three Point Function'',  
Nucl.\ Phys.\ {\bf B622} (2002) 309.
% [arXiv:hep-th/0110244]. 

\bibitem{b-on-sos} 
I.~Kostov, B.~Ponsot, D.~Serban,
%``Boundary Liouville Theory and 2D Quantum Gravity'', 
Nucl.Phys. B683 (2004) 309.% [arXiv:hep-th/0307189].
\bibitem{b-on-sos2}
J.~E.~Bourgine, K.~Hosomichi and I.~Kostov,
%``Boundary transitions of the O(n) model on a dynamical lattice,''
Nucl.\ Phys.\  B {\bf 832}, 462 (2010); %[arXiv:0910.1581 [hep-th]].
J.-E. Bourgine and K. Hosomichi, 
%``Boundary operators in the O(n) and RSOS matrix models",  
JHEP {\bf 0901}:009 (2009).

\bibitem{bloop} 
I.~Kostov, 
%`` Boundary Correlators in 2D Quantum Gravity: Liouville versus Discrete Approach'',
Nucl.Phys. B658 (2003) 397;
% [arXiv:hep-th/0208034];
%``Boundary Loop Models and 2D Quantum Gravity'''
J.\ Stat.\ Mech.\ 0708:P08023 (2007).
% [arXiv:hep-th/0703221].
\bibitem{bloop2}
J.\ Jacobsen, H.\ Saleur, 
Nucl.\ Phys.\ {\bf B788} (2008) 137.

%\cite{Martinec:1991ht}
\bibitem{Martinec:1991ht}
  E.~J.~Martinec, G.~W.~Moore and N.~Seiberg,
  %``Boundary operators in 2-D gravity,''
  Phys.\ Lett.\  B {\bf 263}, 190 (1991).
  %%CITATION = PHLTA,B263,190;%%

\bibitem{b-two-matrix}
K.~Hosomichi, 
%``Minimal Open Strings", 
JHEP {\bf 0806}:029 (2008).
% [arXiv:0804.4721].

\bibitem{MS-lecture}
G.\ Moore and N.\ Seiberg,
``Lectures on RCFT'',
Physics, Geometry and Topology 
(Plenum, New York, 1990).

\bibitem{Banks} 
M.~R.~Douglas,
%``STRINGS IN LESS THAN ONE-DIMENSION AND THE GENERALIZED 
  % K-D-V HIERARCHIES,''
Phys.\ Lett.\  B {\bf 238} (1990) 176;  
T.\ Banks, M.\ Douglas, N.\ Seiberg and S.\ Shenker,
%``Microscopic and macroscopic loops in non-perturbative 
%two dimensional gravity'',
Nucl.~Phys.~{\bf B238} (1990) 279.

\bibitem{BBR} 
G.~Bonelli, L.~Bonora and A.~Ricco,
Phys.~Lett.~{\bf B637} (2006) 310. 

%%%%%%%%%%%%%%%%%%%%%%%%%%%%
 
\end{thebibliography}
\end{document}